\documentclass[10pt]{revtex4}
\usepackage{amsthm}
\usepackage{amssymb}
\usepackage{amsmath}
\usepackage{graphicx}
\def\l{\label}

\def\Th{\Theta}

\def\sig{\sigma}
\def\om{\omega}
\def\udot{\dot{u}}
\def\nab{\nabla}
\def\3nab{\tilde{\nabla}}

\def\lgl{\langle}
\def\rgl{\rangle}
\def\la {\langle}
\def\ra {\rangle}
\def\nn{\nonumber}
\def\c{\mbox{curl}}

\def\hsp5{\hspace{5mm}}
\newcommand{\sfrac}[2]{{\textstyle{#1\over#2}}}
\def\case#1/#2{\textstyle\frac{#1}{#2}}

\def\be {\begin{equation}}
\def\ee {\end{equation}}
\def\ber {\begin{eqnarray}}
\def\eer {\end{eqnarray}}
\def\bea {\begin{eqnarray}}
\def\eea {\end{eqnarray}}

\def\bc {\begin{center}}
\def\ec {\end{center}}
\newcommand{\hs}{\,-\,}

\begin{document}
%%%%%%%%%%%%%%%%%%%%%%%%%%%%%%%%%%%%%%%%%%%%%%%%%%%%%%%%%%%%%%

\title{The evolution of tensor perturbations in scalar-tensor theories of gravity}

\author{Sante Carloni$^{1}$,  Peter K. S. Dunsby$^{1,2}$}

\address{$1.$ Department of Mathematics and Applied Mathematics, \\
University of Cape Town, 7701 Rondebosch, South Africa,}
\address{$2.$ South African Astronomical Observatory, Observatory 7925,
Cape Town, South Africa.}

\begin{abstract}
The evolution equations for tensor perturbations in a generic scalar
tensor theory of gravity are presented. Exact solution are given for
a specific class of theories and
Friedmann-Lema\^{i}tre-Robertson-Walker backgrounds. In these cases
it is shown that, although the evolution of tensor models depends on
the choice of parameters of the theory, no amplification is possible
if the gravitational interaction is attractive.
\end{abstract}

\maketitle

%%%%%%%%%%%%%%%%%%%%%%%%%%%%%%%%%%%%%%%%%%%%%%
\section{Introduction}
%%%%%%%%%%%%%%%%%%%%%%%%%%%%%%%%%%%%%%%%%%%%%%

In spite of the enormous technological problems connected with their
detection, gravitational waves (GW) could become a very important
future source of data in cosmology. Cosmological GW are produced at
very early stages of the evolution of the universe and decouple from
the cosmic fluid at very early times. Consequently they carry
information about  the conditions of the early universe that are not
accessible via the electromagnetic spectrum. One of the most
important applications of a future detection of primordial GW is the
possibility of constraining models of inflation
\cite{Grishchuk:1974ny}.

Even if GW are decoupled from the cosmic fluid their presence still
influences some features of the observable universe. In particular,
a GW background will produce a signature that can be found in the
anisotropies  \cite{Rubakov:1982df}  and polarization
\cite{Hu:1997hv} of the Cosmic Microwaves Background (CMB). This,
together with the extraordinary improvement in the sensitivity of CMB
measurements, opens the possibility of obtaining  important information
about  GW in an indirect way,  even though conventional GW detectors are not
yet full operational.

The study of  cosmological GW via the CMB is also very important
in investigations of alternative theories of gravity. In fact, it is well known
\cite{Will} that the features of GW in GR are rather peculiar and the
detection of any deviation from this behavior would be a genuine
proof of the break down of standard GR.

For this reason it is important to develop a theory of gravitational
waves (or equivalently of tensor perturbations) for alternative
theories of gravity. These problems have been widely studied  using
standard metric-based approaches \cite{Wagoner}, but as yet a fully
covariant and gauge-invariant approach, based on the 1+3 formalism
\cite{EB,EllisCovariant} has not yet  been developed. The aim of
this paper is to show that this method leads to  a clear,
mathematically well defined description of the evolution of tensor
perturbations that, unlike metric based methods is easily
generalized to non--standard gravity.  In order to illustrate the
power of this approach, we will apply it to one of the most studied
extensions of the Einstein theory: Scalar Tensor Gravity. This study
extends a previous work on the covariant gauge invariant scalar
perturbations in this framework \cite{SantePertScTn}.

The paper is organized as follows. In section II we will briefly review
the 1+3 covariant approach in a general setting. In section III we
derive equations that govern the evolution of  tensor perturbations for a
general imperfect fluid. In section IV we adapt these equations the study of
scalar tensor gravity and apply them  to a specific example. Finally, conclusions
are given in section V.

Unless otherwise specified, natural units ($\hbar=c=k_{B}=8\pi G=1$)
will be used throughout this paper, Latin indices run from 0 to 3.
The symbol $\nabla$ represents the usual covariant derivative and
$\partial$ corresponds to partial differentiation. We use the
$-,+,+,+$ signature and the Riemann tensor is defined by
\begin{equation}
R^{a}{}_{bcd}=W^a{}_{bd,c}-W^a{}_{bc,d}+ W^e{}_{bd}W^a{}_{ce}-
W^f{}_{bc}W^a{}_{df}\;,
\end{equation}
where the $W^a{}_{bd}$ is the usual Christoffel symbol
(i.e. symmetric in the lower indices), defined by
\begin{equation}
W^a{}_{bd}=\sfrac{1}{2}g^{ae}
\left(g_{be,d}+g_{ed,b}-g_{bd,e}\right)\;.
\end{equation}
The Ricci tensor is obtained by contracting the {\em first} and the
{\em third} indices
\begin{equation}\label{Ricci}
R_{ab}=g^{cd}R_{acbd}\;.
\end{equation}
Finally the Hilbert--Einstein action in presence of matter is
defined by
\begin{equation}
{\cal A}=\int d x^{4} \sqrt{-g}\left[\sfrac{1}{2}R+ L_{m}\right]\;.
\end{equation}
%%%%%%%%%%%%%%%%%%%%%%%%%%%%%%%%%%%%%%%%%%%%%%
\section{1+3 Covariant approach to cosmology}
%%%%%%%%%%%%%%%%%%%%%%%%%%%%%%%%%%%%%%%%%%%%%%
The starting point (and the corner stone) of our analysis is the 1+3
covariant approach to cosmology \cite{EllisCovariant}. This approach
consists of deriving  a set of first order differential equations and constraints
 for some suitable, geometrically well defined quantities (the {\em 1+3 equations})
 that are completely equivalent to the Einstein field equations. This has the advantage
of simplifying the analysis of general spacetimes which can be
foliated as a set of three dimensional (spacelike) surfaces. In the
following we give a very brief introduction to the parts of this
formalism used in this paper.
%%%%%%%%%%%%%%%%%%%%%%%%%%%%%%%%%%%%%%%%%%%%%%%
\subsection{Kinematics}
%%%%%%%%%%%%%%%%%%%%%%%%%%%%%%%%%%%%%%%%%%%%%%%
In order to derive the 1+3 equations we have to choose a set of
observers i.e. a 4-velocity field $u^a$. This choice depends
strictly on the theory of gravity that we are treating. In this
section we give the set of equations for a general velocity
field and then specialize them to the case of scalar-tensor gravity
in section IV.

Given the velocity $u^a$, we can define the projection tensor into
the tangent 3-spaces orthogonal to the flow vector:
\begin{equation}
h_{ab}  \equiv g_{ab}+u_au_b\; \Rightarrow h^a{}_b
h^b{}_c=h^a{}_c\;, ~h_{ab}u^b=0
\end{equation}
and the kinematical quantities can be obtained by splitting the covariant
derivative of $u_a$ into its irreducible
parts:
\begin{equation}
\nab_b u_a=\3nab_b u_a-A_a u_b\;, ~~~\3nab_b u_a=\sfrac{1}{3}\Theta
h_{ab} +\sigma_{ab}+\omega_{ab}\;, \label{eq:dec}
\end{equation}
where $\3nab_a$ is the spatially totally projected covariant
derivative operator orthogonal to $u^a$, $A_a = \dot{u}_a$ is the
acceleration ($a_bu^b=0$), $\Theta$ is the expansion parameter,
$\sigma_{ab}$ the shear ($\sigma_{ab} =\sigma_{(ab)}$, $\sigma^a{}_a
= \sigma_{ab}u^b=0$) and $\omega_{ab}$ is the vorticity
($\omega_{ab} =\omega_{[ab]}$, $\omega_{ab}u^b=0)$
\footnote{Following the standard convection we will indicate the
symmetrization over two index of a tensor with round brackets and
the antisymmetrization with square ones}.

In the $u^a$ frame, the {\em Weyl or conformal curvature tensor}
$C_{abcd}$ can be split into its electric ($E_{ab}$) and Magnetic  ($H_{ab}$)
respectively:
\bea E_{ab} = C_{acbd}\,u^c\,u^d & \Rightarrow &
E^a{}_a = 0 \ , ~E_{ab} = E_{(ab)} \ , ~E_{ab}\,u^b = 0 \ , \\
H_{ab} = {\sfrac12}\,\eta_{ade}\,C^{de}{}_{bc}\,u^{c} & \Rightarrow
& H^a{}_a = 0 \ , ~H_{ab} = H_{(ab)} \ , ~H_{ab}\,u^b = 0 \ . \eea
In what follows we will use orthogonal projections of vectors
and orthogonally projected symmetric trace-free part of tensors.
They are defined as follows:
\be v^{\la a\ra} = h^{a}{}_{b}\,v^{b} \ , \hsp5 T^{\la ab\ra} = [\
h^{(a}{}_{c}\,h^{b)}{}_{d} - \sfrac{1}{3}\,h^{ab}\,h_{cd}\ ]\,T^{cd}
\ . \ee
Angle brackets may also be used to denote orthogonal
projections of covariant time derivatives along $u^{a}$:
\be \dot{v}{}^{\la a\ra} = h^{a}{}_{b}\,\dot{v}{}^{b} \ , \hsp5
\dot{T}{}^{\la ab\ra} = [\ h^{(a}{}_{c}\,h^{b)}{}_{d} -
\sfrac{1}{3}\,h^{ab}\,h_{cd}\ ]\,\dot{T}{}^{cd} \ . \ee
%%%%%%%%%%%%%%%%%%%%%%%%%%%%%%%%%%%%%%%%%%%%%
\subsection{Thermodynamics}
%%%%%%%%%%%%%%%%%%%%%%%%%%%%%%%%%%%%%%%%%%%%%
The velocity field $u^{a}$ and the projection tensor $h_{ab}$ allow one
to split a generic stress-energy tensor $T_{a b}$ as follows:
\begin{equation}
T_{ab}=\mu u_a u_b +p
h_{ab}+q_au_b+q_bu_a+\pi_{ab}\;, \label{eq:pf}
\end{equation}
where  $\mu$ is the energy density and $p$  is the pressure of the
fluid, $q_a$ represents the energy flux, $\pi_{ab}$ is the
anisotropic pressure and
\bea \hsp5 q_a\,u^a = 0 \ , ~\pi^a{}_a = 0
\ , ~\pi_{ab} = \pi_{(ab)} \ , ~\pi_{ab}\,u^b = 0 \ ;. \nonumber \eea

In a general fluid the pressure, energy density and  entropy are related to
each other by an equation of state $p=p(\mu,s)$. A fluid is considered {\em
perfect} if $q_a$ and $\pi_{ab}$ vanish, and {\em barotropic} if
the entropy is a constant i.e. the equation of state reduces to $p=p(\mu)$.
%%%%%%%%%%%%%%%%%%%%%%%%%%%%%%%%%%%%%%%%%%%%%%
\subsection{Evolution Equations}
%%%%%%%%%%%%%%%%%%%%%%%%%%%%%%%%%%%%%%%%%%%%%%
Writing the Ricci and the Bianchi identities in terms of the $1+3$
variables defined above, we obtain a set of evolution equations \footnote{Here the `curl' is
defined as $(\c\,X)^{ab} = \eta^{cd\lgl a}\,\3nab_{c}X^{b\rgl}\!_{d}$.} :
\be \l{eq:ray} \dot{\Th} - \3nab_{a}\udot^{a} = -
\,\sfrac{1}{3}\,\Th^{2} + (\udot_{a}\udot^{a}) - 2\,\sig^{2} +
2\,\om^{2} - \sfrac{1}{2}\,(\mu+3p) + \Lambda \ , \ee
\be \l{eq:omdot} \dot{\om}^{\lgl a\rgl} - \sfrac{1}{2}\,\eta^{abc}\,
\3nab_{b}\udot_{c} = - \,\sfrac{2}{3}\,\Th\,\om^{a} +
\sig^{a}\!_{b}\,\om^{b} \ ; ,\ee
\bea \l{eq:sigdot} \dot{\sig}^{\lgl ab\rgl} - \3nab{}^{\lgl
a}\udot^{b\rgl} = - \,\sfrac{2}{3}\,\Th\,\sig^{ab} + \udot^{\lgl
a}\, \udot^{b\rgl} - \sig^{\lgl a}\!_{c}\,\sig^{b\rgl c} - \om^{\lgl
a}\,\om^{b\rgl} - (E^{ab}-\sfrac{1}{2}\,\pi^{ab}) \ , \eea
\bea (\dot{E}^{\lgl ab\rgl}+\sfrac{1}{2}\,\dot{\pi}^{\lgl ab\rgl}) -
(\c\,H)^{ab} + \sfrac{1}{2}\,\3nab^{\lgl a}q^{b\rgl} & = & -
\,\sfrac{1}{2}\,(\mu+p)\,\sig^{ab}
- \Th\,(E^{ab}+\sfrac{1}{6}\,\pi^{ab}) \\
& &  + \ 3\,\sig^{\lgl a}\!_{c}\,(E^{b\rgl c}
-\sfrac{1}{6}\,\pi^{b\rgl c}) - \udot^{\lgl a}\,q^{b\rgl}
\nonumber \\
& &  + \ \eta^{cd\lgl a}\,[\ 2\,\udot_{c}\,H^{b\rgl}\!_{d} +
\om_{c}\,(E^{b\rgl}\!_{d}+\sfrac{1}{2}\,\pi^{b\rgl}\!_{d})\ ] \ ,
\nonumber \eea
\bea \dot{H}^{\lgl ab\rgl} + (\c\,E)^{ab} -\sfrac12(\c\,\pi)^{ab} &
= & - \,\Th\,H^{ab} + 3\,\sig^{\lgl a}\!_{c}\,H^{b\rgl c}
+ \sfrac{3}{2}\, \om^{\lgl a}\,q^{b\rgl} \\
& & \hsp5 - \ \eta^{cd\lgl a}\,[\ 2\,\udot_{c}\,E^{b\rgl}\!_{d} -
\sfrac{1}{2}\,\sig^{b\rgl}\!_{c}\,q_{d} - \om_{c}\,H^{b\rgl}\!_{d}\
] \ , \nonumber \eea
\be \l{eq:cons1} \dot{\mu} + \3nab_{a}q^{a} = - \,\Th\,(\mu+p) -
2\,(\udot_{a}q^{a}) - (\sig^{a}\!_{b}\pi^{b}\!_{a})\;, \ee
\be \l{eq:cons2} \dot{q}^{\lgl a\rgl} + \3nab^{a}p +
\3nab_{b}\pi^{ab} = - \,\sfrac{4}{3}\,\Th\,q^{a} -
\sig^{a}\!_{b}\,q^{b} - (\mu+p)\,\udot^{a} - \udot_{b}\,\pi^{ab} -
\eta^{abc}\,\om_{b}\,q_{c} \ , \ee
and a set of constraints
\bea \l{eq:divE}  \3nab_{b}(E^{ab}+\sfrac{1}{2}\,\pi^{ab}) -
\sfrac{1}{3}\,\3nab^{a}\mu + \sfrac{1}{3}\,\Th\,q^{a} -
\sfrac{1}{2}\,\sig^{a}\!_{b}\,q^{b} - 3\,\om_{b}\,H^{ab} - \
\eta^{abc}\,[\ \sig_{bd}\,H^{d}\!_{c} -
\sfrac{3}{2}\,\om_{b}\,q_{c}\ ]  = 0\;, \eea
\bea \l{eq:divH}\3nab_{b}H^{ab} + (\mu+p)\,\om^{a} +
3\,\om_{b}\,(E^{ab}-\sfrac{1}{6}\,\pi^{ab})+ \ \eta^{abc}\,[\
\sfrac{1}{2}\,\3nab_{b}q_{c} + \sig_{bd}\,(E^{d}\!_{c}
+\sfrac{1}{2}\,\pi^{d}\!_{c})\ ] &=& 0\;, \eea
\be\l{eq:onu} \3nab_{b}\sig^{ab} - \sfrac{2}{3}\,\3nab^{a}\Th +
\eta^{abc}\,[\ \3nab_{b}\om_{c} + 2\,\udot_{b}\,\om_{c}\ ] + q^{a}
=0 \;, \ee
\be   \3nab_{a}\om^{a} - (\udot_{a}\om^{a}) = 0\;, \ee
\be \l{hconstr}  H^{ab} + 2\,\udot^{\lgl a}\, \om^{b\rgl} +
\3nab^{\lgl a}\om^{b\rgl} - (\c\,\sig)^{ab} = 0\;, \ee
that are completely equivalent to the Einstein equations. It is from these equations that we
derive the general evolution equations for the tensor perturbations in scalar-tensor gravity.

%The Riemann and the Ricci tensor can also be easily expressed in
%terms of these quantities (see \cite{EllisCovariant}) as well as
%their equivalent on the three dimensional hypersurfaces.  Of
%particular interest for us will be the Ricci scalar for the 3-spaces
%orthogonal to the velocity field \be \l{eq:3R} \tilde{R} = 2\,\mu -
%\sfrac{2}{3}\,\Th^{2} + 2\,\sig^{2} + 2\,\Lambda \;. \ee

%%%%%%%%%%%%%%%%%%%%%%%%%%%%%%%%%%%%%%%%%%
\section{Perturbations Equations}
%%%%%%%%%%%%%%%%%%%%%%%%%%%%%%%%%%%%%%%%%%
\subsection{The Background}
%%%%%%%%%%%%%%%%%%%%%%%%%%%%%%%%%%%%%%%%%%
The equations presented in the previous section hold in any
spacetime we wish  to analyze. However, in what follows we will
focus on the class of spacetimes that do not differ too much from a
Friedmann-Lema\^{i}tre-Robertson-Walker (FLRW) model. The reason is
that current observations suggest that the universe appears to
deviate only slightly from homogeneity and isotropy. We can define a
FLRW spacetimes in terms of the variables above.  Homogeneity and
isotropy imply:
 \begin{equation}
 \sigma =\omega= 0\;, ~~\3nab_a f=0\;,
 \label{eq:rwcond1}
 \end{equation}
 where $f$ is any scalar quantity; in particular
 \begin{equation}
 \3nab_a\mu =\3nab_a p=0 ~~\Rightarrow~~, ~\dot{u}_a = 0  \;.
 \label{eq:rwcond3}
 \end{equation}
It follows that  the governing equations for this background are
 \begin{equation}
\dot{\Theta}+\sfrac{1}{3} \Theta^2+\sfrac{1}{2}\left(\mu+3p\right)=
0\;, \label{eq:rayback}
\end{equation}
\begin{equation}
\tilde{R} = 2\left[- \sfrac{1}{3} \Theta^2+\mu\right]\;,
\end{equation}
\begin{equation} \label{cons2}
    \dot{\mu_{\phi}}+\Theta\left(\mu+p\right)=0\;.
\end{equation}
Now in order to describe small deviations from a FLRW spacetime
we simply take all the quantities that are zero in the
background as being first order, and retain in the equations
(\ref{eq:ray}-\ref{hconstr}) only the terms that are linear in these quantities, i.e.
we drop all second order terms. This procedure corresponds to the {\em
linearization} in the 1+3 covariant approach and it greatly
simplifies the system (\ref{eq:ray}-\ref{hconstr}). In particular,
the scalar vector and tensor parts of the perturbations are
decoupled, so that we are able to treat them separately. In what
follows we will focus only on the tensor perturbations.
%%%%%%%%%%%%%%%%%%%%%%%%%%%%%%%%%%%%%%%%%%%
\subsection{Covariant decomposition of the 1+3 variables}
%%%%%%%%%%%%%%%%%%%%%%%%%%%%%%%%%%%%%%%%%%%
Even if the geometrical nature of the kinematical 1+3 variables are
known from their definition, the same cannot be said for the
thermodynamic variables and the terms that contain derivatives. In
the first case, this is because they depend strictly on the details
of the energy momentum tensor of the fluid we are dealing with and
in the second case, the index carried by the derivative might, in
principle, change the geometrical nature of the quantity. For this
reason, we have to define a general procedure that allows us to
recognize the nature of these new quantities or split them in their
irreducible parts. Giving a completely general procedure is a highly
non-trivial task. Fortunately, however, for vectors and trace-free
symmetric tensor, which describe the key quantities in the covariant
approach, this task is relatively easy and can be summarized as
follows.

In the case of a vector field the decomposition procedure is based
on the Helmholtz's theorem \cite{Sommerfeld}. This theorem says that
a continuous vector field $V_a$ that vanishes (together with its
first derivative) on the boundaries of a manifold $\mathcal{M}$ can
be decomposed, modulus a constant, into an irrotational field
$\tilde{V}_a$ and a solenoidal field $\bar{V}_a$
\footnote{Helmholtz's theorem is given in a modern and more
generalized form in  the Kodaira-Hodge-De Rham decomposition theorem
\cite{KHdR}.}:
\begin{equation}
    V_a= \bar{V}_a + \hat{V}_a=\eta^{abc}\3nab_{b}\bar{V}_{c}+\3nab^a \hat{v}\,;,
\end{equation}
where
\begin{equation}
    \3nab^a \bar{V}_a =0\;, \qquad  \eta^{abc}\3nab_{b}\hat{V}_{c}=0\;.
\end{equation}
With a little more work we can generalize this decomposition to any
symmetric tensor that vanish on the boundary of $\mathcal{M}$
\cite{Stewart} obtaining:
\begin{equation}
    W_{ab}=
     \bar{W}_{ab}+\hat{W}_{ab}+W^{*}_{ab}= \bar{W}_{ab} +
    \3nab_a\bar{W}_b +  \3nab_a\3nab_b W^{*}\;,
\end{equation}
where
\begin{equation}
    \3nab^a \bar{W}_{ab} =0\;, \qquad (\c\,\hat{W})_{ab} =0\;,
    \qquad (\c\,W^{*})_{ab}=0
\end{equation}
and both of these decompositions are unique. Now we can define
scalars, vectors or tensors as quantities that transform like
scalars, solenoidal vectors or symmetric tensors, or are obtained
from them using the $h_{ab}$ or $\3nab_{a}$ operators.

%The same can be
%done with solenoidal vectors, transverse traceless tensors and any possible
%higher rank quantity.

In what follows we will consider only purely tensor perturbations
which are characterized by the fact that vector and scalar contributions
 to the 1+3 variables and equations vanish, i.e.
\begin{equation} \label{tensor1}
    f,\; \3nab^a f,\;\3nab^a \3nab^b f=0,\quad \forall\; f\;\;
    \mbox{(scalar)},
\end{equation}
and
\begin{equation}\label{tensor2}
    \bar{V}_a,\; \3nab_a \bar{V}_b =0,\quad \forall\; \bar{V}_a\;\;
\mbox{(solenoidal vector)}\;.
\end{equation}
%%%%%%%%%%%%%%%%%%%%%%%%%%%%%%%%%%%%%%%%%%%
\subsection{Tensor perturbation equations}% general form of $\pi$
%%%%%%%%%%%%%%%%%%%%%%%%%%%%%%%%%%%%%%%%%%%
Focusing only on tensor perturbations,  i.e. applying
(\ref{tensor1}) and (\ref{tensor2}), the evolution equations
(\ref{eq:ray})-(\ref{hconstr}) reduce to
\begin{equation}\label{eqsigma}
    \dot{\sigma} _{ab} + \frac{2}{3}\,\Theta \,\sigma _{ab}
  + E_{ab} - \frac{1}{2} \pi _{ab}=0\;,
\end{equation}
\begin{equation}\label{eqmag}
   \dot{H}_{ab}+
   H_{ab}\,\Theta+(\c\,E)_{ab} - \frac{1}{2}(\c\,\pi)_{ab} = 0\;,
\end{equation}
\begin{equation}\label{eqelect}
   \dot{E}_{ab}+E_{ab}\,\Theta -
  (\c\,H)_{ab}  +
   \frac{1}{2}\left( p + \mu \right) \,\sigma_{ab}
   + \frac{1}{6}\Theta\,\pi_{ab} +
   \frac{1}{2} \dot{\pi}_{ab}=0\;,
\end{equation}
together with the conditions
\begin{equation}\label{constraints}
    \3nab_{b}H^{ab}=0\;,\quad \3nab_{b}E^{ab}=0\;,\quad H_{ab}=(\c\,\sigma)_{ab}\;.
\end{equation}
Taking the time derivative of the above equations we obtain
\begin{equation}\label{eq2ordSigma}
 \ddot{\sigma}_
  {ab}-\3nab^{2}\sigma  +
  \frac{5}{3}\, \Theta \, \dot{\sigma}_{ ab} +
  \left(\frac{1}{9}\, {\Theta }^2 + \frac{1}{6}\, \left(1 - 9\,w \right)\, \mu  \right)\, \sigma_ {ab}
   =-\left(\dot{\pi}_{ab}+ \frac{2}{3}\, \Theta \,
   \pi_{ab}\right)\;,
\end{equation}
\begin{equation}\label{eq2ordMag}
\ddot{H}_{ab} - \3nab^{2}H_{ab} +
  \frac{7}{3}\, \Theta\, \dot{H}_{ab} + \frac{2}{3}\, \left({\Theta }^2 - 3\, w\,
\mu\right)\, H_{ab}
     = (\c\, \pi)^{\cdot}_{ab}+\frac{2 }{3}\, \Theta\, (\c\, \
\pi)_{ab}\;,
\end{equation}
\begin{eqnarray}\label{eq2ordElect}
 \nn  \ddot{E}_{ab}&-&\3nab^{2}E_{ab}+ \frac{7}{3}\,\Theta\,\dot{E}_{ab} +
  \frac{2}{3}\,
     \left( {\Theta}^2 - 3\,w\,\mu \right) {E}_{ab}  -
  \frac{1 }{6}\Theta\,\mu\,\left( 1 + w \right) \,
     \left( 1 + 3\,c_{s}^{2}\right)\,\sigma _{ab} \\ &=& -
     \left[\frac{1}{2}\ddot{\pi}_{ab}-\frac{1}{2}\3nab^{2}\pi_{ab}
       +\frac{5}{6}\,\Theta\,\dot{\pi}_{ab} +
  \frac{1}{3}\left( {\Theta }^2 - \mu \right) \,\pi_{ab}\right]\;,
\end{eqnarray}
where $c_{s}^{2}$ is the sound speed of the fluid
and we have used the Raychaudhuri equation
(\ref{eq:ray}), the energy conservation equation (\ref{cons2}) and
the commutator identity
\begin{equation}\label{}
   (\c\,\dot{X})_{ab}=(\c\,X)^{\cdot}_{ab}+
  \frac{1}{3}(\c\,X)\,\Theta\;.
\end{equation}
These equations generalize the  tensor perturbation equations for an
imperfect fluid that were derived in \cite{Challinor} (note the different signature!).

Once the form of the anisotropic pressure has been determined the
(\ref{eq2ordSigma}-\ref{eq2ordElect})  can be solved to give the
evolution of tensor perturbations. As already noted in
\cite{PeterTensor} the presence of a term that contains the shear in
the (\ref{eq2ordElect}) makes this equation effectively third  order,
so that it is not possible to write down a closed wave equation
for $E_{ab}$. If $\pi_{ab}=0$, it is easy to show that for consistency,
the solution for this field must also satisfy a wave equation because the
shear is a solution of a wave equation and equation (\ref{eqsigma}) holds.
This will also be the case here because in our case $\pi_{ab}\propto
\sigma_{ab}$ and so the anisotropic pressure will also  satisfy a wave equation.
%%%%%%%%%%%%%%%%%%%%%%%%%%%%%%%%%%%%%%%%%%%%%%
\subsection{Harmonic analysis}
%%%%%%%%%%%%%%%%%%%%%%%%%%%%%%%%%%%%%%%%%%%%%%
Following standard harmonic analysis, equations (\ref{eq2ordSigma}) and (\ref{eq2ordMag})
may be reduced to ordinary differential equations. It is standard \cite{bi:BDE} to use trace-free
symmetric tensor eigenfunctions of the spatial  the Laplace-Beltrami operator defined by:
\begin{eqnarray}\label{eq:harmonic}
  \3nab^{2}Q_{ab} = -\frac{k^{2}}{a^{2}}Q_{ab}\;,
\end{eqnarray}
where $k=2\pi a/\lambda$ is the wavenumber and $\dot{Q}_{ab}=0$. Developing $\sigma_{ab}$ and $H_{a b}$ in terms of the
$Q_{ab}$, (\ref{eq2ordSigma}) and (\ref{eq2ordMag}) reduce to
\begin{equation}\label{eq2ordSigmaHarm}
 \ddot{\sigma}^{(k)} +
  \frac{5}{3}\, \Theta \, \dot{\sigma}^{(k)} +
  \left(\frac{1}{9}\, {\Theta }^2 + \frac{1}{6}\, \left(1 - 9\,w \right)\, \mu -\frac{k^{2}}{a^{2}}  \right)\, \sigma^{(k)}_ {ab}
   =\dot{\pi}^{(k)}+ \frac{2}{3}\, \Theta \,
   \pi^{(k)}\;,
\end{equation}
\begin{equation}\label{eq2ordMagHarm}
\ddot{H}^{(k)} +
  \frac{7}{3}\, \Theta\, \dot{H}^{(k)} + \frac{2}{3}\, \left({\Theta }^2 - 3\, w\,
\mu-\frac{k^{2}}{a^{2}}\right)\, H^{(k)}
     = (\c\, \dot{\pi})^{(k)}+\frac{2 }{3}\, \Theta\, (\c\, \
\pi)^{(k)}\;,
\end{equation}
and equation (\ref{eqelect}) reads
\begin{equation}\label{eq2ordElectHarmLWL}
E^{(k)}=- \dot{\sigma}^{(k)} - \frac{2}{3}\,\Theta \,\sigma ^{(k)} +
\frac{1}{2} \pi ^{(k)}\,.
\end{equation}
%If we consider only wavelengths much bigger than the Hubble radius
%$\lambda\gg H^{-1}$, the equations (\ref{eq:harmonic}) imply that
%all the Laplacians terms can be neglected and the
%(\ref{eq2ordSigma}) and (\ref{eq2ordMag}) can be written as
%\begin{equation}\label{eq2ordSigmaHarmLWL}
% \ddot{\sigma}^{(k)} +
%  \frac{5}{3}\, \Theta \, \dot{\sigma}^{(k)} +
%  \left[\frac{1}{9}\, {\Theta }^2 + \frac{1}{6}\, \left(1 - 9\,w \right)\, \mu \right]\,
%\sigma^{(k)} =\left(\dot{\pi}^{(k)}+ \frac{2}{3}\, \Theta \,
%   \pi^{(k)}\right)\;,
%\end{equation}
%\begin{equation}\label{eq2ordMagHarmLWL}
%\ddot{H}^{(k)} +
%  \frac{7}{3}\, \Theta\, \dot{H}^{(k)} + \frac{2}{3}\, \left({\Theta }^2 - 3\, w\,
%\mu\right)\, H^{(k)}
%     = [(\c\, \pi)^{\cdot}]^{(k)}+\frac{2 }{3}\, \Theta\, (\c\, \
%\pi)^{(k)}\;,
%\end{equation}
%while (\ref{eq2ordElectHarm}) remains unchanged.In the following we
%will use both (\ref{eq2ordSigmaHarm}-\ref{eq2ordElectHarmLWL}) and
%(\ref{eq2ordElectHarmLWL}-\ref{eq2ordMagHarmLWL}).

%%%%%%%%%%%%%%%%%%%%%%%%%%%%%%%%%%%%%%%%%%%%%
\section{Tensor perturbation in Scalar Tensor gravity}
%%%%%%%%%%%%%%%%%%%%%%%%%%%%%%%%%%%%%%%%%%%%%
\subsection{Scalar Tensor gravity}
%%%%%%%%%%%%%%%%%%%%%%%%%%%%%%%%%%%%%%%%%%%%%
In Scalar-Tensor theories of gravity the inclusion of Mach's
principle implies the introduction of a scalar field $\phi$
non-minimally coupled to the geometry \cite{bi:BransDicke}. In
particular, this field is associated with the gravitational
``constant" $G$ and leads to it varying in time. Like many of the
extensions of GR, scalar-tensor theories appear in many fundamental
schemes and have been proposed as models for dark energy because
their cosmology naturally leads to the phenomenon of {\em cosmic
acceleration}, which is the characteristic footprint of dark energy
\cite{ScTnDarkEnergy,Claudio}.

The  most general  action for Scalar-Tensor Theories of gravity is
given by (conventions as in Wald \cite{bi:wald})
\begin{equation}\label{eq:actionScTn}
\mathcal{A}=\int d x^{4}\sqrt{-g}\left[\sfrac{1}{2}F(\phi)R
-\sfrac{1}{2}\nab_a\phi\nab^a\phi -V(\phi)+\mathcal{L}_m \right]\;,
\end{equation}
where $V(\phi)$ is a general (effective) potential expressing the
self interaction of the scalar field and $\mathcal{L}_m$ represents
the matter contribution.

Varying the action with respect to the metric gives the
gravitational field equations:
\begin{eqnarray}
&& \nn F(\phi)G_{ab}= F(\phi)\left(R_{ab}-\sfrac{1}{2}\,g_{ab}
R\right)=\\&&=T _{ab}^{m}+\nab_a\phi\nab_b\phi-g_{ab}
\left(\sfrac{1}{2}\nab_c\phi\nab^c\phi+V(\phi)\right)
+\left(\nab_b\nab_a - g_{ab}\nab_c\nab^c\right)F(\phi)\;,
\label{eq:einstScTn}
\end{eqnarray}
while the variation with respect to the field $\phi$ gives the
curved spacetime version of the Klein\hs Gordon equation
\begin{equation}
\nab_a\nab^a\phi+\sfrac{1}{2}F'(\phi) R -V'(\phi)=0\;, \label{eq:KG}
\end{equation}
where the prime indicates a derivative with respect to $\phi$. Both
these equations reduce to the standard equations for GR and a
minimally coupled scalar field when $F(\phi)=1$.

Equation (\ref{eq:einstScTn}) can be recast in the form:
\begin{equation}
\label{eq:einstScTneff}
 G_{ab}=T^{TOT}_{ab}=\frac{ T_{ab}^{m}}{F(\phi)}+T^{(eff)}_{ab}\,,
 \end{equation}
where $T^{(eff)}_{ab}$ is defined as
\begin{eqnarray}\label{eq:TenergymomentuEff}
\nonumber &&T^{(eff)}_{ab}=T_{ab}^{\phi}=
\frac{1}{F(\phi)}\left[\nab_a\phi\nab_b\phi-g_{ab}
\left(\sfrac{1}{2}\nab_c\phi\nab^c\phi+V(\phi)\right) +\nab_b\nab_a
F(\phi)- g_{ab}\nab_c\nab^cF(\phi)\right]\;, \label{eq:semt}
\end{eqnarray}
and  an effective gravitational coupling $G_{eff}= F(\phi)^{-1}$
appears. Provided that $\phi_{,a} \neq 0$, equation (\ref{eq:KG})
also follows from the conservation equations
\begin{equation}
\nab^bT_{ab}^{\phi}=0\;.
 \label{eq:cons}
\end{equation}

%%%%%%%%%%%%%%%%%%%%%%%%%%%%%%%%%%%%%%%%%
\subsection{Perturbation equations in Scalar-Tensor gravity}
%%%%%%%%%%%%%%%%%%%%%%%%%%%%%%%%%%%%%%%%%

At this point we are ready to derive the equations for the evolution of
tensor perturbations in scalar-tensor gravity. The key point we will use here
is that, as we have seen, the field equations of these models can be
recast in the form (\ref{eq:einstScTneff}), where together with matter,
an effective fluid  with energy momentum tensor
$T^{\phi}_{ab}$ appears. Following  the prescription given in
section II, it is then relatively easy to generalize
(\ref{eq2ordSigma}-\ref{eq2ordElect}) to this case.

A crucial point in this procedure is the choice of frame: since
(\ref{eq:einstScTneff}) describes a two fluids system, it is not
immediately obvious which choice of $u^a$ is the more convenient. It
turns out that if we assume standard matter to be a perfect fluid, the
best choice of frame is one comoving with the effective fluid. In
this  frame, standard matter is in general imperfect, i.e. it
acquires a frame-induced heat flux and anisotropic pressure.
However, it can be shown \cite{bi:KingEllis}, that for a perfect
fluid only the frame--induced heat flux is first order, i.e. standard
matter does not produce any frame--induced first order tensor
contributions. Consequently, in the effective fluid frame we avoid
any frame--induced contributions to both the effective fluid
thermodynamic quantities and standard matter. We therefore
obtain the simplest possible form for the  total anisotropic pressure
appearing in the R.H.S. of equations (\ref{eq2ordSigma}-\ref{eq2ordElect}).

In order to define the effective fluid frame, we assume that the
{\it momentum density} $\nab^a\phi$ is {\it timelike}
\cite{bi:peterscalar,bi:madsen,SantePertScTn}:
\begin{equation}
\nab_a\phi\nab^a\phi<0\;, \label{eq:ass1}
\end{equation}
in an open region $A$ of spacetime and we
choose the 4-velocity to be
\begin{equation}
 u^a\equiv - \frac{\nab^a\phi}{\psi}\;,~~~u^a u_a=-1\;,~~~\psi
\equiv \dot{\phi} = u^a\nab_a\phi = (-\nab_a\phi
\nab^a\phi)^{1/2}\;. \label{eq:u}
\end{equation}
This implies
\begin{equation}
\3nab_a\phi=0, \label{eq:gfi}
\end{equation}
and
\begin{equation}
\omega_{ab}=-h_a{}^c
h_b{}^d\nab_{[d}\left(\frac{1}{\psi}\nab_{c]}\phi\right)=0\;,
\end{equation}
i.e. our foliation selects vorticity free spacelike hypersurfaces in
which $\phi=const$.

Using the choice (\ref{eq:u}), the tensor $T^{TOT}_{ab}$ can be
decomposed  as follows  \cite{SantePertScTn}:
\begin{equation}\label{eq:ed}
\mu_{TOT} =\mu_m +\frac{1}{F(\phi)}\left[\sfrac{1}{2}\psi^2 +V(\phi)
-\Theta\dot{F}(\phi)\right]\;,
\end{equation}
\begin{equation}\label{eq:p}
p_{TOT}=p_m +\frac{1}{F(\phi)}\left[\sfrac{1}{2}\psi^2 -V(\phi)
+\left(\ddot{F}(\phi)+\sfrac{2}{3}\Theta\dot{F}(\phi)\right)\right]\;,
\end{equation}
\begin{equation}
q_a^{TOT}=q^m_a +\frac{\dot{F}(\phi)}{F(\phi)}\; \dot{u}_a
\label{eq:qflux}\;,
\end{equation}
\begin{equation}
\pi_{ab}^{TOT}=\pi_{ab}^{m}-\frac{\dot{F}(\phi)}{F(\phi)}\;\sigma_{ab}\;.
\label{eq:anisp}
\end{equation}
As anticipated, $\pi_{ab}^{m}$ does not affect our tensor
perturbations equations, so that the total anisotropic pressure in
the frame $u_a$ is proportional to the shear. This allows us to
rewrite (\ref{eq2ordSigma}-\ref{eq2ordElect}) as
\begin{eqnarray}\label{eqPertTensScTnSigma}
 \nn\ddot{\sigma}^{(k)}+ \left( \frac{5}{3}\,\Theta  +
 \frac{\dot{F}}{F} \right) \,\dot{\sigma}^{(k)}\  -
\left(\frac{\dot{F}^2}{{F}^2} -\frac{2}{3}\,\Theta
\,\frac{\dot{F}}{F} -\frac{\ddot{F}}{F} -\frac{1}{9}\,{\Theta }^2 -
\frac{1}{6}\,\left( 1 -9\,w \right) \,\mu
-\frac{k^{2}}{a^{2}}\right)  \sigma^{(k)}=0
\end{eqnarray}
\begin{eqnarray}\label{eqPertTensScTnMag}
% \nonumber to remove numbering (before each equation)
  \nn\ddot{H}^{(k)}+
  \left( \frac{7}{3}\,\Theta  +
     \frac{\dot{F}}{F} \right) \,
   \dot{H}^{(k)}  -
     \left( 2 w\,\mu- \frac{2}{3}{\Theta }^2\,
          + \frac{\dot{F}^2 }{{F}^2}-\,\Theta \,\frac{\dot{F}}{F} -
          \frac{\ddot{F}}{F}-\frac{k^{2}}{a^{2}}  \right)  H^{(k)}=0
\end{eqnarray}
\begin{eqnarray}\label{eqPertTensScTnElect}
% \nonumber to remove numbering (before each equation)
 \nn\ddot{E}^{(k)}&+&
  \frac{7}{3}\,\Theta \,\dot{E}^{(k)}   +
  \left( \frac{2}{3}\,{\Theta }^2 -
     2\,w\,\mu  - \frac{3}{2}\,\frac{\dot{F}^2}{{F}^2}
       \frac{\ddot{F}}{F}-\frac{k^{2}}{a^{2}} \right)E^{(k)}\nn \\
  &-& \left( \frac{1}{6}\,\mu
\,\Theta \,\left( 1 + w \right) \,\left( 1 + 3\, c_s^2  \right)-
\frac{1}{12}\, \left( 5 - 9\,w \right) \,\mu
\,\frac{\dot{F}}{F}+\frac{5}{18}\,{\Theta
}^2\,\frac{\dot{F}}{F}  \right.\nn\\
&-& \left.\frac{1}{6} \Theta \,\frac{\dot{F}^2}{F^2}+
\frac{9}{4}\,\frac{\dot{F}^3}{{F}^3} -
          \frac{5}{2}\,\frac{\dot{F}}{F}\frac{\ddot{F}}{F}+
          \frac{1}{6}\,\Theta \,\frac{\ddot{F}}{F}
          + \frac{1}{2}\,\frac{\dddot{F}}{F}- \frac{\ddot{F}}{F}\frac{k^{2}}{a^{2}}\right)\sigma^{(k)}=0\;.
\end{eqnarray}
These equations describe the evolution of tensor perturbations for a
general scalar tensor theory of gravity and reduce to the ones
presented in \cite{PeterTensor,Challinor} in the GR limit
$(F\rightarrow 1)$.

In the next subsection, we will apply these equations to a specific
example of coupling, potential and background.

%%%%%%%%%%%%%%%%%%%%%%%%%%%%%%%%%%%%%%%%
\subsection{The case $F(\phi)=-\xi\phi^2$, $V(\phi)=\lambda\phi^{p}$ in vacuum}
\begin{figure}
  \includegraphics[width=13cm]{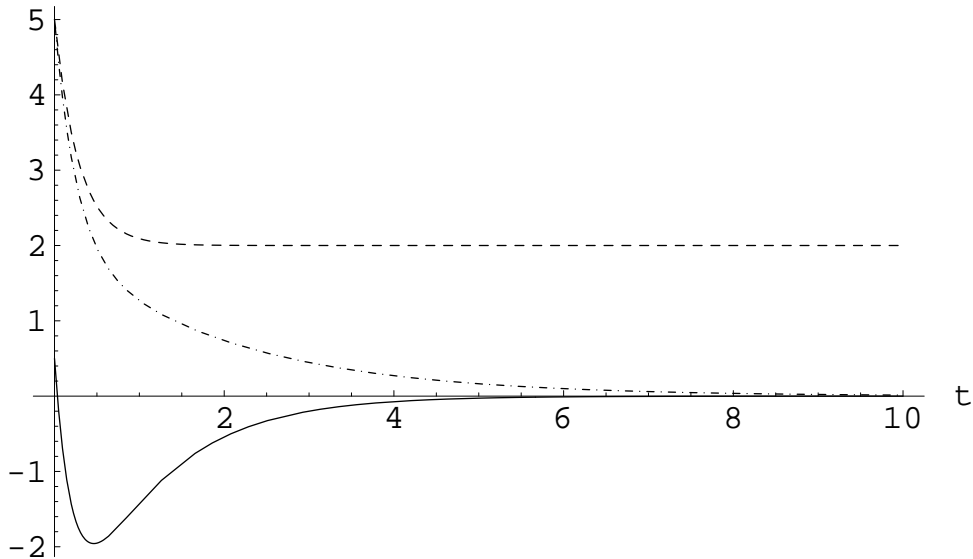}\\
\caption{Plot of the solutions of (\ref{LWLsolScTn}) for a scalar
tensor theory of gravity in vacuum with $F(\phi)=-\xi\phi^2$,
$V(\phi)=\lambda\phi^{p}$, the background (\ref{eq:NoethSol}),
$k=10$ and $r=-1$. The dashed represents $\sigma^{(k)}$,  the
dot-dashed curve represents $H^{(k)}$ and the solid curve represents
$E^{(k)}$ }\label{fig:PlotScTnLWLs1}
\end{figure}

%%%%%%%%%%%%%%%%%%%%%%%%%%%%%%%%%%%%%%%%
As an example, let us consider the class of scalar tensor theories given by
\begin{equation}\label{couplingF}
 F(\phi) = - \xi (r) \phi^2,\quad  V(\phi) = \lambda  \phi^{2 p(r)},
\end{equation}
where $\xi$ and $ p$  depend on a single free parameter $r$, via
\begin{equation}
 \xi (r) = \frac{(2r+3)^2}{12 (r+1)(r+2)} \quad ; \quad
 p(r) = \frac{3(1+r)}{3+2r}\;.
 \end{equation}
Using the Noether symmetry approach \cite{bi:salvreview} it can be
shown that this class of theories admit the following
exact solution in a vacuum:
\begin{equation}\label{eq:NoethSol}
a(t) = \alpha \,t^n \quad ; \quad \phi(t) = \beta\, t^m ;\quad
K=0\;,
\end{equation}
with
\begin{equation}
n(r) = \frac{2 r^2 + 9 r + 6}{r(r+3)} \quad ;
 \quad m(r) = - \frac{2r^2 + 9 r + 9}{r(r+3)}\;.
\end{equation}
This represents a specific case of a general exact analytic
solution for the background equations
\cite{SantePertScTn,bi:marino,bi:salvreview}. Depending on the value
of $r$, the (\ref{eq:NoethSol}) can represent power law inflation
behaviour ($n(r)>1$) or a Friedmann-like phase ($0<n(r)<1$)
\footnote{This solution  can represent, of course, also a
contraction, but in the following we will consider only the
expanding cases.} and its character is essentially related to the
choice of the form of the potential (via the choice of $r$). The
parameter $ \alpha $ is free while $ \beta $ is linked to $ \lambda
$ and $ r $ through
$$
\lambda = \frac{(6+r)(3+2r)^2}{8r(3+r)^2(2+3r+r^2)}
\beta^{-\frac{2r}{3+2r}}.
$$
It turns out to be simpler to leave $\beta$ free and derive $\lambda$, but
of course the real free parameter is this last one. Let us write
down the time dependence of some important quantities. The expansion
parameter is
$$
\Theta(t) = 3\frac{n(r)}{t}\;.
$$
The effective energy density is given by
$$
\mu(t) = \frac{3(6+9r+2r^2)^2}{r^2(3+r)^2}\frac{1}{t^2}\;,
$$
which is a non negative function of $r$. Finally the effective barotropic
index is
$$
w=- \frac{2}{3}  -
  \frac{2 + r}
   {6 + r\,\left( 9 + 2\,r \right) },
$$
which is constant. Substituting in the equations
(\ref{eqPertTensScTnSigma}) and (\ref{eqPertTensScTnMag}) we obtain
\begin{eqnarray}\label{eqExScTnSigHarm}
\ddot{\sigma}^{(k)} + \frac{3\,\left( 4 + r \right) \,\left( 1 +
2\,r \right)}{r\,\left( 3 + r \right) \,t} \,\dot{\sigma}^{(k)} +
\left[\frac{3\,\left( 1 + r \right) \,\left( 4 + r \right) \,\left(
9 + 2\,r \right)}{r\,{\left( 3 + r
\right)}^2\,t^2}+\frac{k}{t^{2n(r)}}\right]\,\sigma^{(k)}=0\;,
\end{eqnarray}
\begin{eqnarray}\label{eqExScTnMAgHarm}
\ddot{H}^{(k)}&+&\frac{24 + 5\,r\,\left( 9 + 2\,r \right)
}{r\,\left( 3 + r \right) \,t}\,\dot{H}^{(k)}+ \left[\frac{2 (s (2
s+9)+3) (s (5
   s+24)+18)}{s^2 (s+3)^2 t^2}+\frac{k}{t^{2n(r)}} \right]\;H^{(k)} =0\;.
\end{eqnarray}
%for short wavelengths and
%\begin{eqnarray}\label{eqExScTnSigHarm}
%\ddot{\sigma}^{(k)} + \frac{3\,\left( 4 + r \right) \,\left( 1 +
%2\,r \right)}{r\,\left( 3 + r \right) \,t} \,\dot{\sigma}^{(k)} +
%\left[\frac{3\,\left( 1 + r \right) \,\left( 4 + r \right) \,\left(
%9 + 2\,r \right)}{r\,{\left( 3 + r
%\right)}^2\,t^2}\right]\,\sigma^{(k)}=0
%\end{eqnarray}
%\begin{eqnarray}\label{eqExScTnMAgHarm}
%\nn&&\ddot{H}^{(k)}-\frac{24 + 5\,r\,\left( 9 + 2\,r \right)
%}{r\,\left( 3 + r \right) \,t}\,\dot{H}^{(k)}+\\&&+
%\left[\frac{1}{t^2}\left(28 + \frac{24}{r^2} + \frac{66}{r}+
%  \frac{24}{3 + r} + \frac{12}{{\left( 3 + r \right) }^2}\right) \right]\;H^{(k)} =0
%\end{eqnarray}
The two equations above give the behaviour of the shear and the
gravitomagnetic field. The gravitoelectric field can be obtained
from the solution of the equations above and (\ref{eqsigma})
\begin{equation}
\nn E^{(k)}=-\dot{\sigma}^{(k)}-\frac{\left( 3 + 9\,r + 2\,r^2
\right) } {r\,\left( 3 + r \right) \,t}\;\sigma^{(k)}\;.
\end{equation}
In the long wavelength limit ($k=0$), the above equations admit the
solution:
\begin{eqnarray}\label{LWLsolScTn}
\sigma^{(k)} &=& C_1 t^{3+\frac{2}{r+3}+\frac{4}{r}}+C_2 t^{\frac{r}{r+3}-3}\;, \\
  H^{(k)} &=& C_1 t^{-5-\frac{3}{r+3}-\frac{6}{r}}+C_2 t^{-4-\frac{4}{r+3}-\frac{2}{r}}\;, \\
  E^{(k)} &=& C_1 t^{-4-\frac{2}{r+3}-\frac{4}{r}} -C_2 t^{-\frac{3 (r+4)}{r+3}}\;,
\end{eqnarray}
which are plotted for $r=-1$  in Figure \ref{fig:PlotScTnLWLs1}.
Note that  the exponents in the above solution depend on $r$  so
that there are values of $r$ for which the above solution present
growing modes i.e. the gravitational waves are amplified on
superhorizon scale. Since also the gravitational constant depends on
the parameter $r$, we are interested in  values of this parameter
for which we have growing modes {\em and} an attractive
gravitational interaction. Unfortunately, it can be easily shown
that such modes are incompatible with attractive gravitational
interaction i.e. the values of $r$ for which the exponents in the
solutions above are positive imply a negative effective
gravitational constant. This means that in this class of theories
and background only cosmological models driven by a repulsive
gravitational constant induce an amplification of gravitational
waves.
%This is possible in the following cases:
%
%\begin{equation}\label{}
%\begin{array}{cc}
% \sigma^{(k)}\;\; \mbox{growing} \Rightarrow &  -\frac{9}{2}<r<-3, -1<r<0 \\
%  H^{(k)} \;\; \mbox{growing} \Rightarrow  & \left\{\begin{array}{c}
%                                                        \frac{3}{5} \left(-4+\sqrt{6}\right)<r<0 \\
%                                                        \frac{1}{4}  \left(-9+\sqrt{57}\right)<r<0
%                                                    \end{array}\right.\\
%  E^{(k)}\;\; \mbox{growing} \Rightarrow  & -4<r<-3,\frac{1}{4}
%   \left(-9+\sqrt{33}\right)<r<0
%\end{array}
%  % \left\{\left\{-\frac{9}{2}<r<-3,-4<r<-3,-1<r<0\right\},\left\{\frac{3}{5}
%%   \left(-4-\sqrt{6}\right)<r<-3,\frac{3}{5}
%%   \left(-4+\sqrt{6}\right)<r<0,\frac{1}{4}
%%   \left(-9-\sqrt{57}\right)<r<-3,\frac{1}{4}
%%   \left(-9+\sqrt{57}\right)<r<0\right\},\left\{-4<r<-3,\frac{1}{4}
%%   \left(-9-\sqrt{33}\right)<r<-3,\frac{1}{4}
%%   \left(-9+\sqrt{33}\right)<r<0\right\}\right\}\right)
%\end{equation}

\begin{figure}
  \includegraphics[width=13cm]{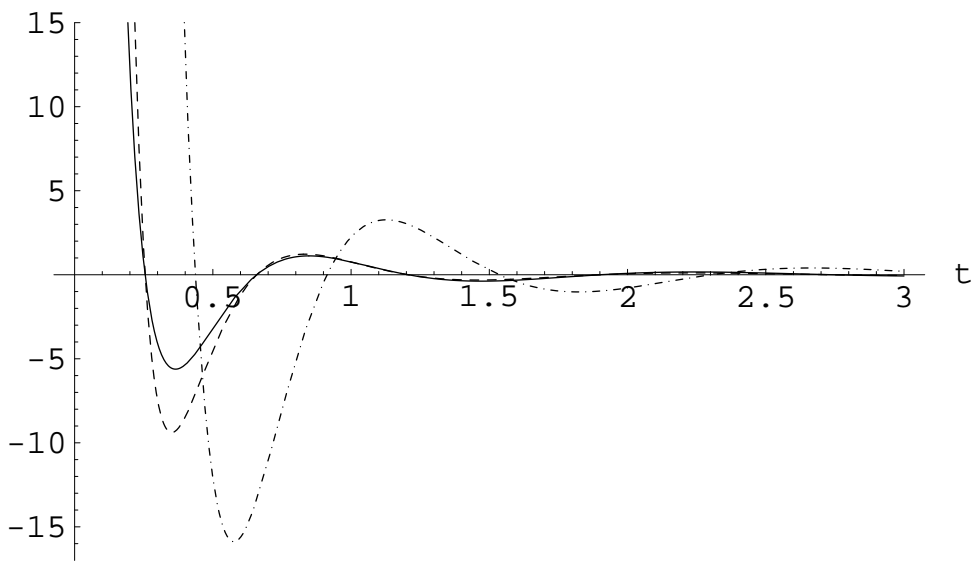}\\
\caption{Plot of the solutions of (\ref{SWsolScTn}) for a scalar
tensor theory of gravity in vacuum with $F(\phi)=-\xi\phi^2$,
$V(\phi)=\lambda\phi^{p}$, the background (\ref{eq:NoethSol}),
$k=10$ and $r=-1$. The solid curve represents $\sigma^{(k)}$,  the
dashed curve represents $H^{(k)}$ and the dot-dashed curve
represents $E^{(k)}$ }\label{fig:PlotScTnSWs1}
\end{figure}

In the general case, the solutions for (\ref{eqExScTnSigHarm}) and
(\ref{eqExScTnMAgHarm}) involve Bessel functions of first and second
kind $J$ and $Y$. We obtain:
\begin{eqnarray}\label{SWsolScTn}
  \nonumber\sigma^{(k)} &=& t^{-\frac{5 (r+4)}{2(r+3)}-\frac{2}{r}}\left[C_1\; J\left(\frac{1}{2}+\frac{3}{r
    (r+6)+6},\;\frac{k \;r\;(r+3)\; t^{-\frac{1}{r+3}-1-\frac{2}{r}}}{(r
    (r+6)+6) \sqrt{\alpha}}\right)\right.\\
 &&+\left. C_2\;
    Y\left(\frac{1}{2}+\frac{3}{r
    (r+6)+6},\frac{k \;r\;(r+3)\; t^{-\frac{1}{r+3}-1-\frac{2}{r}}}{(r (r+6)+6)
    \sqrt{\alpha}}\right)\right]\;,\\
 \nonumber H^{(k)} &=&t^{-\frac{9}{2}-\frac{4}{r}-\frac{7}{2
    (r+3)}} \left[C_1\; J\left(\frac{1}{2}+\frac{3}{r
    (r+6)+6},\;\frac{k \;r\;(r+3)\;
    t^{-\frac{1}{r+3}-1-\frac{2}{r}}}{(r (r+6)+6)
    \sqrt{\alpha}}\right)\right.\\
 &&+\left. C_2\;Y\left(\frac{1}{2}+\frac{3}{r (r+6)+6},\frac{k
    \;r\;(r+3)\; t^{-\frac{1}{r+3}-1-\frac{2}{r}}}{(r (r+6)+6)
    \sqrt{\alpha}}\right)\right]\;, \\
\nonumber E^{(k)} &=& t^{-\frac{7}{2}-\frac{2}{r}-\frac{5}{2
    (r+3)}} \left[C_1\;
    J\left(\frac{1}{2}+\frac{3}{r (r+6)+6},\;\frac{k \;r\;(r+3)\;
    t^{-\frac{1}{r+3}-1-\frac{2}{r}}}{(r (r+6)+6)
    \sqrt{\alpha}}\right)\right.\\
  \nonumber &&+\left. C_2\;
    Y\left(\frac{1}{2}+\frac{3}{r (r+6)+6},\frac{k \;r\;(r+3)\;
    t^{-\frac{1}{r+3}-1-\frac{2}{r}}}{(r (r+6)+6)
    \sqrt{\alpha}}\right)\right]\\
 \nonumber &&+
    \frac{m}{\sqrt{\alpha}}t^{-\frac{9}{2}-\frac{4}{r}-\frac{7}{2
    (r+3)}}\left[C_1\; J\left(\frac{3}{2}+\frac{3}{r (r+6)+6},\;\frac{k
    \;r\;(r+3)\; t^{-\frac{1}{r+3}-1-\frac{2}{r}}}{(r (r+6)+6)
    \sqrt{\alpha}}\right)\right.\\
 &&+\left. C_2\;
    Y\left(\frac{3}{2}+\frac{3}{r (r+6)+6},\frac{k \;r\;(r+3)\;
    t^{-\frac{1}{r+3}-1-\frac{2}{r}}}{(r (r+6)+6)
    \sqrt{\alpha}}\right)\right]\;.
\end{eqnarray}
The form of these solutions are similar to the one found in GR and
this was expected  from the structure of the equations. A plot of
these solutions for $r=-1$ is given in Figure
\ref{fig:PlotScTnSWs1}. As in the long wavelength limit also here
growing modes are possible in principle but not compatible with
attractive gravitational interaction so that also in short
wavelengths only negative $G_{eff}$ cosmological models induce an
amplification of gravitational waves.

%\begin{figure}
%  \includegraphics[width=13cm]{PlotScTnLWLs417.eps}\\
%  \caption{Plot of the long wavelength solutions of (\ref{SWsolScTn}) for
%a scalar tensor theory of gravity in vacuum with
%$F(\phi)=-\xi\phi^2$, $V(\phi)=\lambda\phi^{p}$, the background
%(\ref{eq:NoethSol}), $k=10$ and $r=-4.17$. The dashed represents
%$\sigma^{(k)}$,  the dot-dashed curve represents $H^{(k)}$ and the
%solid curve represents  $E^{(k)}$ }\label{fig:PlotScTnLWLs417}
%\end{figure}
%\begin{figure}
%  \includegraphics[width=13cm]{PlotsScTnSWs417.eps}\\
%  \caption{Plot of the long wavelength solutions of (\ref{LWLsolScTn}) for
%a scalar tensor theory of gravity in vacuum with
%$F(\phi)=-\xi\phi^2$, $V(\phi)=\lambda\phi^{p}$, the background
%(\ref{eq:NoethSol}), $k=10$ and $r=-4.17$. The dashed curve
%represents $\sigma^{(k)}$,  the solid curve represents $H^{(k)}$ and
%the dot-dashed curve represents  $E^{(k)}$
%}\label{fig:PlotScTnSWs417}
%\end{figure}

%%%%%%%%%%%%%%%%%%%%%%%%%%%%%%%%%%%%%%%%%%%%%%%%%
\section{Conclusions}
%%%%%%%%%%%%%%%%%%%%%%%%%%%%%%%%%%%%%%%%%%%%%%%%%
In this paper we have derived the evolution of the tensor
perturbations in a general scalar-tensor theory of gravity. These
equations have been obtained using the covariant gauge invariant
approach to perturbation theory. The key element of this approach
is the fact that the Scalar-Tensor field equations can be recast in
a form similar to the GR field equations, where the energy-momentum is
described standard matter plus an effective fluid connected to the
non--minimally coupled scalar field appears. In this way,
equations for the evolution of the tensorial perturbations can
be derived in a straightforward way.

To illustrate the method, we analyzed the the evolution of
tensor perturbations of a class of theories and backgrounds
obtained by the Noether symmetry approach \cite{bi:salvreview}. Our
analysis reveals that, although the details of the evolution of the gravitational waves
depend on the specific values of the parameters, no growing modes are compatible
with an attractive gravitational interaction.

%In the treatment of the scalar perturbations of the same class of
%models and theories \cite{ScTnPertSan}, we found that there is a
%value of the parameter for which a non--minimally coupled scalar
%field mimics a perfect fluid with zero barotropic index
%($\omega=0$). The above analysis reveals that although $\phi$  and
%dust behave in the same way in terms of the scalar perturbations
%they have a different behavior in terms of the tensor ones. This is
%particularly evident in the short wavelengths in which the shear
%perturbations of the non--minimally coupled scalar field grows
%instead of decreasing. Since the time evolution of the tensor
%perturbation has a direct influence on the features of the
%polarization and tensorial spectrum of the Cosmic Microwave
%Background, such difference offers a new way to distinguish between
%dust and minimally coupled scalar field via observations.
%\section*{References}

\end{document}